\documentclass[12pt,a4paper,reprint,aps,pra]{revtex4-1}
\usepackage[english]{babel}
\usepackage
{graphicx}
\usepackage{epstopdf}
\usepackage{amsmath}
\usepackage{latexsym}
\usepackage{mathrsfs}
\usepackage{dcolumn}
\usepackage{color}
\usepackage{bm}
\usepackage{units,nicefrac}
\usepackage{natbib}

\begin{document}

\title{Morphology of Critically-Sized Crystalline Nuclei \\ at Shear-Induced Crystal Nucleation in Amorphous Solid}

\author{Bulat N. Galimzyanov}
\email{bulatgnmail@gmail.com}
\affiliation{Kazan Federal University, 420008 Kazan, Russia}

\author{Anatolii V. Mokshin}
\affiliation{Kazan Federal University, 420008 Kazan, Russia}

\date{\today}
\begin{abstract}
In this work we study morphological characteristics of the
critically-sized crystalline nuclei at initial stage of the
shear-induced crystallization of a model single-component
amorphous (glassy) system. These
characteristics are estimated quantitatively through statistical
treatment of the non-equilibrium molecular dynamics simulation
results for the system under steady shear at various (fixed)
values of the shear rate $\dot\gamma$ and at different temperatures. It is found that the
sheared glassy system is crystallized through nucleation
mechanism. From analysis of a time-dependent trajectories of the
largest crystalline nuclei, the critical size $n_{c}$ and the
nucleation time $\tau_{c}$ were defined. It is shown that the
critically-sized nuclei in the system are oriented within the shear-gradient $xy$-plane at moderate and high shear
rates; and a tilt angle of the oriented nuclei depends on the shear rate. At
extremely high shear rates and at shear deformation of the system
more than $60$\%, the tilt angle of the nuclei tends to take
the value $\simeq45^\circ$ respective to the shear direction. We found
that this feature depends weakly on the temperature. Asphericity
of the nucleus shape increases with increasing shear rate,  that
is verified  by increasing value of the asphericity parameter and by
the contour of the pair distribution function  calculated for the
particles of the critically-sized nuclei. The critical size
increases with increasing shear rate according to the power-law,
$n_{c}\propto(\dot\gamma\tau_{c})^{1/3}$, whereas the shape of the
critically-sized nucleus changes from spherical to the
elongated ellipsoidal. We found that the $n_{c}$-dependencies
of the nuclei deformation parameter evaluated for the system at
different temperatures and shear rates are collapsed into unified
master-curve.
\end{abstract}

\maketitle

\section{Introduction}

Due to disordered structure, the amorphous materials (metallic
glasses, polymeric and colloidal amorphous solids) have unique
mechanical and physical properties, which allow one to find their
practical applications in photonics, medicine
and electronics~\cite{Greer_1995,Lu_Liu_2003,Scelsi_McLeish_2009,Mokshin_Barrat_2010,Berthier_2011,Ediger_Harrowell_2012,Janeschitz-Kriegl_2013}.
Furthermore, the microscopic structure of disordered systems under
different mechanical deformations (compression, shear etc.) is of a
special interest~\cite{Lowen_2001}. In particular, understanding of
physical mechanisms of the steady shear influence on microscopic
structure of the systems could provide a possibility to develop
practical tools to control the structural ordering
process~\cite{Butler_Harrowell_1995,Haw_Pusey_1998,Evans_Morriss_2008,He_Lowen_2009,Scelsi_McLeish_2009,Dasgupta_Procaccia_2012,Mokshin_Barrat_2010,Fang_Wang_2013}.

A large number of experimental and simulation studies provide
indications that the steady shear applied to metastable disordered
systems (supercooled liquids and amorphous solids) generates a
microscopic structure anisotropy in these systems~\cite{Fang_Wang_2013,Shao_Singer_2015,Mura_Zaccone_2016,Kumaraswamy_Kornfield_2002,Koumakis_Petekidis_2012,Kerrache_2011,Butler_Harrowell_1995,Blaak_PRL_93_2004,Mokshin_Barrat_2010,
Mokshin_Galimzyanov_Barrat_2013,Nosenko_Ivlev_2012}. It is clear
that the anisotropy arisen in the disordered systems under shear
leads to change of morphological characteristics of emerging
crystalline structures. Therefore, the mechanical and rheological properties of the systems can be significantly dependent on the anisotropy. This is confirmed by results of large number of
studies (Refs.~\cite{Keller_Machin_1967,Butler_Harrowell_1995,Viasnoff_2002,Rottler_2003,Wallace_2006,Olmsted_PRL_103_2009,Graham_Olmsted_2010,McIlroya_Olmsted_2017,Mokshin_Barrat_2010,Janeschitz-Kriegl_2013,Koumakis_Petekidis_2016,Roozemond_Peters_2013,Roozemond_Peters_2015}).
So, for example, Kumaraswamy and
co-workers~\cite{Kumaraswamy_Kornfield_2002} have revealed
experimentally that non-spherical
ordered structures oriented along the shear direction occur for the case of polydisperse isotactic
polypropylene melt under intensive shear.
Moreover, they found that
these structures appear as precursors of crystallization at a temperature below the melting point; and
morphological properties of these precursors have a significant
impact on the overall crystallization kinetics. On the basis of
experimental and simulation studies done by Petekidis and
co-workers~\cite{Koumakis_Schofieldc_Petekidis_2008,Koumakis_Petekidis_2012,Koumakis_Petekidis_2016}, it was found that the microscopic structure of
polymeric colloidal glasses under shear changes with increasing
shear rate. This is directly manifested in the shape of the
evaluated density distribution
$g(x,y)$.
Namely, the function $g(x,y)$ characterizing the particles
distribution within the shear-gradient plane starts to change its
shape under applied shear flow. Then, it is reasonably to expect that the observed structural changes of the systems may do impact on their rheological properties. It was found in
Ref.~\cite{Koumakis_Schofieldc_Petekidis_2008} that the ordered
structures forming under shear are not stable thermodynamically,
and they may be characterized by rheological aging. These results
are supplemented by simulation studies. For example, simulation
results of Blaak \emph{et al.} have revealed that the
elongated crystalline nuclei oriented along the shear direction
occur at shear-induced crystallization of colloidal
suspensions~\cite{Blaak_PRL_93_2004}. Here, a tilt angle of the
nucleus increases with increasing shear rate. On the other hand,
Graham and Olmsted found that non-spherical elongated nuclei occur
in sheared polymer melts, and such the shape has been mentioned as
shish-kebab~\cite{Olmsted_PRL_103_2009}. According to results of Ref.~\cite{Olmsted_PRL_103_2009}, the
crystal nucleation may be accelerated by
shear. Note, the similar results were also obtained for the case of crystallization of model glassy
systems at homogeneous
shear~\cite{Mokshin_Barrat_2009,Mokshin_Barrat_2010,Mokshin_Galimzyanov_Barrat_2013}.
It was found in Refs.~\cite{Mokshin_Barrat_2009,Mokshin_Barrat_2010,Mokshin_Galimzyanov_Barrat_2013} that steady shear can both accelerate and suppress crystal nucleation process in the glassy systems.

In spite of large number of studies mentioned above, there are still a lot of unclear points related with influence of a steady homogeneous shear on the morphology of
crystalline structures emerging in amorphous solids~\cite{Mokshin_Galimzyanov_Barrat_2013,Mokshin_Galimzyanov_2015,Shao_Singer_2015,Shrivastav_Horbach_2016}.
For example, the physical mechanisms determining the crystalline
nuclei asphericity at the initial stage of crystallization, when
the size of these nuclei is comparable with a critical size, are
debated~\cite{Blaak_PRL_93_2004,Olmsted_PRL_103_2009,Mokshin_Barrat_2010,Mokshin_Galimzyanov_Barrat_2013}.
According to the classical nucleation
theory~\cite{Kashchiev_Nucleation_2000,Kelton_Greer_2010,Kalikmanov_2012},
only the critically-sized nuclei may demonstrate a stable
growth. Moreover, from experimental point of view it is not so easy to evaluate such morphological characteristics as the critical size and the shape
of the crystalline nuclei in the atomistic and molecular systems
under shear with different
rates~\cite{Koumakis_Schofieldc_Petekidis_2008,Fang_Wang_2013,Wang_Liu_2016}.
This is especially appropriate for the case of deep supercooling
levels, when the temperatures are below the glass transition
temperature $T_g$.

In the present work, we study the microscopic mechanisms of
shear-induced crystallization of a single-component glassy system
at different temperatures. The main attention is paid to study
morphological characteristics of the critically-sized crystalline nuclei in the system under a homogeneous shear with various (fixed) shear rates. These characteristics are considered as
the terms dependent on the critical deformation
$\gamma_{c}\equiv\dot\gamma\tau_{c}$. Here, the dimensionless
quantity $\gamma_{c}$ is the deformation of the system under shear
with the rate $\dot\gamma$ at the time moment
$\tau_{c}\equiv\tau(n_{c})$, i.e. when the critically-sized
$n_{c}$ nucleus appears. In other words, the quantity $\gamma_{c}$
corresponds directly in the time scale to the nucleation time
$\tau_{c}$. We find that the homogeneous shear initiates the
structural ordering in the glassy system through the
crystal nucleation mechanism, where the nuclei with pronounced asphericity of their shape and oriented along the shear direction are formed.
Simulation details and cluster analysis are presented in Section
II. The results are given and discussed in Section III.
Finally, the main results will be summarized in the conclusion.

\section{Simulation Details and Cluster Analysis}

Molecular dynamics simulations are performed for the single-component system, where the interaction between the particles is defined through the short-ranged oscillatory Dzugutov potential (Dz-system)~\cite{Dzugutov_1992,Roth_2005}. The specific shape of the potential allows one to reproduce effectively the effect of electron screening for the ion-ion interaction in metals. In addition, the system with such the potential is capable to generate a stable amorphous state~\cite{Mokshin_Barrat_PRE_2008, Mokshin_Barrat_2009, Mokshin_Barrat_2010}.

\begin{figure}[ht]
\centering
\includegraphics[width=0.9\linewidth]{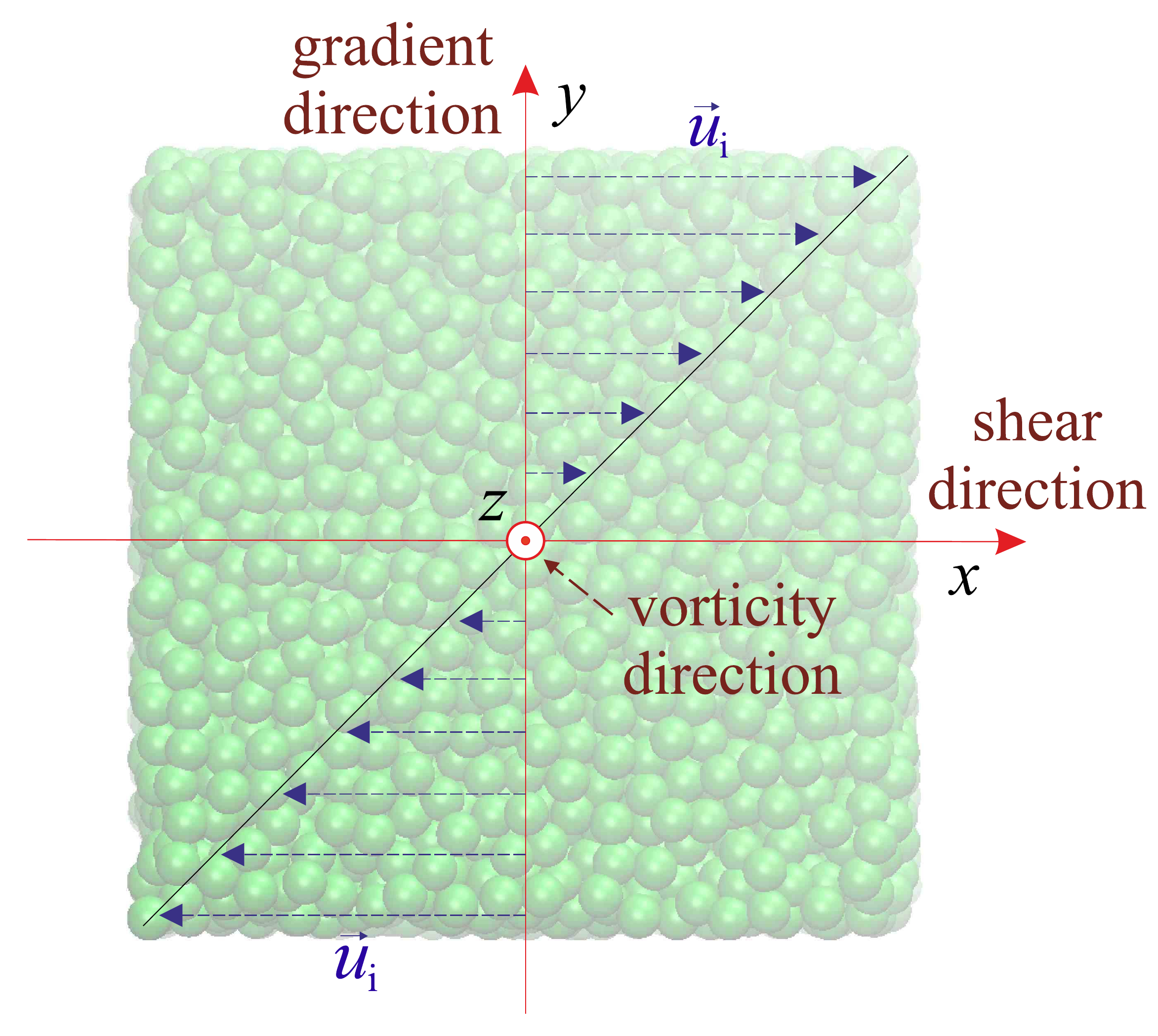}
\caption{(color online) Schematic representation of the homogeneous shear realization with a linear velocity profile: green spheres show the particles of a system; the blue arrows characterize the shear directions with rate $\vec{u}_{i}$ ($i=1,\,2,...,\,N$). The axis $OX$ is associated with the shear-direction; the axis $OY$ corresponds to the gradient-direction; the axis $OZ$ indicates the vorticity-direction. \label{fig_1}}
\end{figure}

To realize the homogeneous shear, the SLLOD-algorithm is applied~\cite{Evans_Morriss_2008}, where the shear velocity
\begin{equation}
\vec{u}_{i}=\dot{\gamma}y_{i}\vec{e}_{x}\label{eq_shear1}
\end{equation}
is added to the $x$-component of intrinsic (thermal) velocity $\vec{\upsilon}_{i}$ of each particle of the system. As a result, equations of motion are~\cite{Evans_Morriss_2008}:
\begin{subequations}\label{eq_emove}
\begin{equation}
\frac{d\vec{r}_{i}}{dt}=\vec{\upsilon}_{i}+\dot{\gamma}y_{i}\vec{e}_{x}+\zeta\vec{r}_{i},\label{eq_emove1}
\end{equation}
\begin{equation}
\frac{d\vec{\upsilon}_{i}}{dt}=\frac{\vec{F}_{i}}{m}-\dot{\gamma}\upsilon_{yi}\vec{e}_{x}-(\varrho+\zeta)\vec{\upsilon}_{i}.\label{eq_emove2}
\end{equation}
\end{subequations}
Here, $\vec{r}_{i}(x_{i},y_{i},z_{i})$ and
$\vec{\upsilon}_{i}(\upsilon_{xi},\upsilon_{yi},\upsilon_{zi})$ are the position and velocity of the $i$th particle ($i=1,\,2,\,...\,N$, where $N$ is the number of particles); $\varrho$ and $\zeta$ are the parameters of thermostat and barostat, respectively. In the present work, the homogeneous shear is applied with different fixed shear rates: $\dot{\gamma}=0.0001$, $0.0005$, $0.001$, $0.002$, $0.005$, $0.008$, and $0.01\,\tau^{-1}$. Realization of the homogeneous shear is schematically presented in Fig.~\ref{fig_1}.

The considered three-dimensional system consists of $N=6912$ identical particles located within the cubic simulation cell. The Lees-Edwards periodic boundary conditions are applied in both the gradient $y$- and vorticity $z$-directions~\cite{Lees_Edwards_1972}. The ordinary periodic boundary conditions are also applied in the shear $x$-direction. The instantaneous velocities and trajectories of the particles are determined by integration of the equations of motion (\ref{eq_emove}) using the Verlet algorithm with the time step $\Delta t=0.005\,\tau$. All simulations are performed in the isobaric-isothermal ensemble, where the temperature $T$ and the pressure $p$ are controlled by the Nose-Hoover thermostat and barostat, respectively. In the present work, the standard Lennard-Jones units are used: $\sigma$ is the effective particle diameter; $\epsilon$ is the energy unit; $\tau=\sigma\sqrt{m/\epsilon}$ is the time unit, where $m$ is the mass of the particle; the temperature $T$ in units of $\epsilon/k_{B}$, where $k_{B}$ is the Boltzmann constant; the pressure $p$ in units of $\epsilon/\sigma^{3}$; the shear rate $\dot\gamma$ in units of $\tau^{-1}$.

Molecular dynamics simulations are realized as follows. At first, a liquid sample is equilibrated at the temperature $T=2.3\,\epsilon/k_{B}$ and at the pressure $p=14\,\epsilon/\sigma^{3}$, that is above the melting temperature of the system $T_{m}\simeq1.51\,\epsilon/k_{B}$ at considered pressure (the phase diagram of the Dz-system can be found in Ref.~\cite{Roth_2005}). Second, the glassy samples are prepared through rapid cooling of the equilibrated liquid with the cooling rate of $0.001\,\epsilon/(k_{B}\tau)$ to the temperatures $T=0.05$, $0.15$, and $0.5\,\epsilon/k_{B}$, that is much lower than the glass transition temperature $T_{g}\simeq0.65\,\epsilon/k_{B}$ of the system. Thus, one hundred independent glassy samples are prepared for each considered ($p$, $T$) thermodynamic state, that are required to perform a statistical treatment of the simulation results. And, finally, each glassy sample is exposed to a homogeneous shear on the time scale $\sim100\,000$ simulation steps.

To detect the crystalline structures, the cluster analysis based on the estimation of the bond orientational order parameters was applied~\cite{Steinhardt_1983,Reinhardt_Doye_2012}. Namely, following to Steinhardt \emph{et al.}~\cite{Steinhardt_1983}, the local orientational order parameters are defined as
\begin{equation}
q_{l}(i)=\left(\frac{4\pi}{2l+1}\sum_{m=-l}^{l}\left|\frac{1}{n_{b}^{(i)}}\sum_{j=1}^{n_{b}^{(i)}}Y_{lm}(\theta_{ij},\varphi_{ij})\right|^{2}\right)^{1/2},\label{eq_local_bop}
\end{equation}
where
\begin{equation}
l=4,\,6,\,8. \nonumber
\end{equation}
Also we estimate value of the global orientational order parameter as an average of $q_{6}(i)$ (i.e. at $l=6$) over all the particles of the system~\cite{Steinhardt_1983}:
\begin{equation}
Q_{6}=\left(\frac{4\pi}{13}\sum_{m=-6}^{6}\left|\frac{\sum_{i=1}^{N}\sum_{j=1}^{n_{b}^{(i)}}Y_{6m}(\theta_{ij},\varphi_{ij})}{\sum_{i=1}^{N}n_{b}^{(i)}}\right|^{2}\right)^{1/2}.\label{eq_Q6_bop}
\end{equation}
Here, $n_{b}^{(i)}$ is the number of neighbors of the $i$th particle; $Y_{lm}(\theta_{ij},\varphi_{ij})$ are the spherical harmonics;
$\theta_{ij}$ and $\varphi_{ij}$ are the polar and azimuthal angles, respectively. As a rule, even spherical harmonics with the indexes $l=4$, $6$ and $8$ are sufficient to be evaluated recognize presence of the ordered domains in the system. For example, the local orientational order parameters $q_{4}$, $q_{6}$, $q_{8}$ take non-zero values for the particles forming $sc$ (simple cubic), $icos$ (icosahedral), $bcc$ (base-centered cubic), $fcc$ (face-centered cubic), and $hcp$ (hexagonal close-packed) crystalline structures~\cite{Mickel_Mecke_2013}. For a disordered glassy system, the local orientational order parameters obey a normal distribution, whereas the value of the parameter $Q_{6}$ tends to zero for this case ~\cite{Steinhardt_1983,Wolde_Frenkel_1996}. The order parameters $q_i$ and $Q_j$, where $i,\,j = 4, 6, 8,...,$ take unique values for specific crystalline structures, that allows one to identify these structures in  considered system~\cite{Steinhardt_1983}. In particular, according to original normalization of the parameter $Q_6$, the parameter $Q_6$ is $0.575$ for the perfect \emph{fcc} lattice; one has $Q_6=0.485$ for the perfect \emph{hcp} lattice and $Q_6=0.663$ for the \emph{icosahedral} lattice. The smaller values can be due to the presents of defects.

One of the key conditions, by which the particle is identified as entering into an ordered phase, is the so-called coherence condition~\cite{Wolde_Frenkel_1996}
\begin{subequations}\label{eq_bop_condition}
\begin{equation}
\left|\sum_{m=-6}^{6}\vec{q}_{6m}^{*}(i)\cdot\vec{q}_{6m}(j)\right|\geq0.5,\label{eq_bop_condition_1}
\end{equation}
\begin{equation}
\vec{q}_{6m}(i)=\frac{q_{6m}(i)}{\sqrt{\sum_{m=-6}^{6}\left|q_{6m}(i)\right|^{2}}}.\label{eq_bop_condition_2}
\end{equation}
\end{subequations}
Namely, according to the scheme suggested in Ref.~\cite{Wolde_Frenkel_1996}, the pair of particles $i$ and $j$ are considered as ``solid-like'' if the dot-product $\vec{q}_{6}^{*}(i)\cdot\vec{q}_{6}(j)$ exceeds the threshold value $0.5$. In this case, the nearest neighborhood of the $i$th particle must contain seven and more ``solid-like'' particles, for which  condition (\ref{eq_bop_condition}) is satisfied.

For quantitative characterization of the crystalline nuclei shape, we evaluate the asphericity parameter~\cite{Reinhardt_Doye_2012}
\begin{equation}
S_{0}=\left\langle\frac{(I_{xx}-I_{yy})^{2}+(I_{xx}-I_{zz})^{2}+(I_{yy}-I_{zz})^{2}}{2(I_{xx}+I_{yy}+I_{zz})^{2}}\right\rangle,\label{eq_So_parameter}
\end{equation}
where
\begin{equation}
I_{\alpha\beta}=m\sum_{i=1}^{n_{c}}(\vec{r}_{i}^{\,2}\delta_{\alpha\beta}-\vec{r}_{i\alpha}\vec{r}_{i\beta})\label{eq_inertia_tensor}
\end{equation}
are the moments of inertia ($\alpha\beta\in\{x,y,z\}$). Here, $|\vec{r}_{i}|$ is the distance between the nucleus center-of-mass and the position of $i$th ``solid-like'' particle; $\delta_{\alpha\beta}$ is the Kronecker delta. Thus, for the case of a perfect spherical shape, we have $S_{0}=0$; the parameter is $S_{0}\rightarrow1$ for extremely elongated shape.

By means of the cluster analysis, we obtain for each $\alpha$th simulation run the growth trajectories of the largest crystalline nucleus, $n_{\alpha}(t)$. Here the quantity $n_{\alpha}(t)$ indicates that the nucleus of the size $n$ appears at the time $t$ during the $\alpha$th simulation run, where $\alpha=1,2,...,100$. These trajectories extracted from the different simulation runs are treated within the mean-first-passage-time method~\cite{Mokshin_Galimzyanov_2012,Mokshin_Galimzyanov_2015}. According to this method, the curve $\tau(n)$ is defined, which is known as the mean-first-passage-time curve and which characterizes the average time of the first appearance of the largest nucleus with given size $n$ (for details, see Refs.~\cite{Mokshin_Galimzyanov_2012,Mokshin_Galimzyanov_2015}). The critical size $n_c$ and the average nucleation (or waiting) time for the nucleus $\tau_c$, are defined from the analysis of the curve $\tau(n)$ and of the first derivative $\partial\tau(n)/\partial n$, according to the scheme suggested in Ref.~\cite{Mokshin_Galimzyanov_2015}. In the case of activation type processes, the curve $\tau(n)$ characterizes by three regimes: (i) - the first regime is associated with pre-nucleation, where the small values of n correspond to $\tau(n)$ with zero value; (ii) - the second regime, in which the curve $\tau(n)$ has the pronounced non-zero slope, contains information about a nucleation event. Namely, detected from the first derivative $\partial\tau(n)/\partial n$ location of an inflection point in the curve $\tau(n)$ for this regime defines the critical size $n_c$, whereas $\tau(n_c)\equiv\tau_c$ is directly associated with the nucleation time of the critically-sized nucleus; (iii) - the third regime, where the slope of $\tau(n)$ decreases corresponds to growth of the nucleus. Note that in the present work we focus on the characteristics for the largest crystalline nucleus.

\section{Results}

\subsection{Crystalline structures}

\begin{figure}[ht]
\centering
\includegraphics[width=0.9\linewidth]{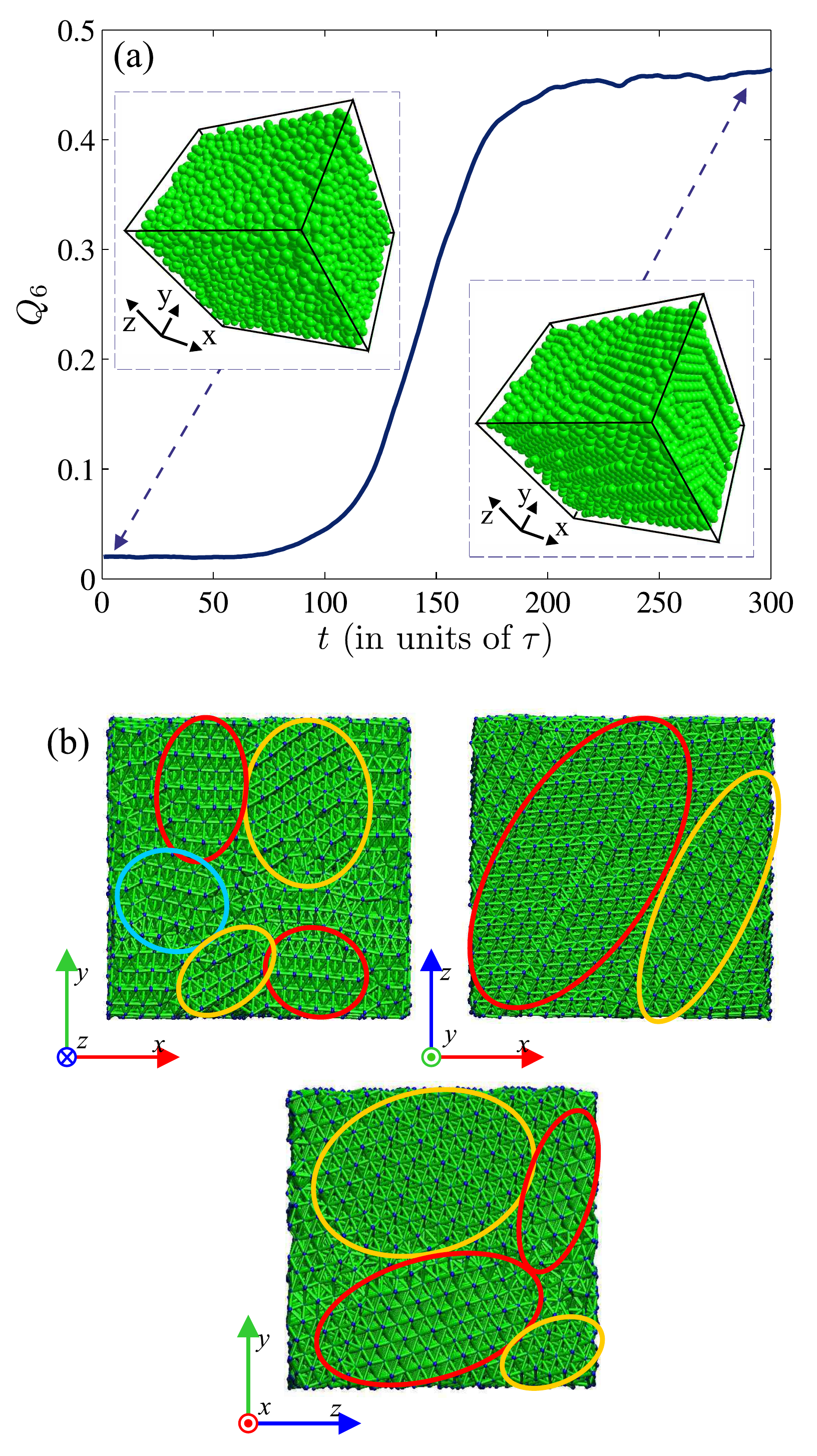}
\caption{(color online) (a) Time-dependent orientational order parameter $Q_{6}(t)$ evaluated for the system at the temperature $T=0.15\,\epsilon/k_{B}$ and at shear rate $\dot\gamma=0.001\,\tau^{-1}$. Insets: instantaneous configurations of a system corresponding to values
$Q_{6}\simeq0.023$ (left inset) and $Q_{6}\simeq0.47$ (right inset). (b) Snapshots of the simulation cell corresponding to the system at the temperature $T=0.15\,\epsilon/k_B$ and the shear rate $\dot{\gamma}=0.001\,\tau^{-1}$, when an ordered state is achieved. Here, the order parameter $Q_6$ is $\sim0.47$. In fact, the system represents a nanocrystalline solid consisted of an assembly of crystallites with disordered orientations, and the degree of order achieved by the system is low in comparison to that of a perfect crystal. \label{fig_2}}
\end{figure}

Information about structural transformations in the glassy system at homogeneous shear is obtained through evaluation of the orientational order parameters $q_{4}$, $q_{6}$, $q_{8}$ and $Q_{6}$. The most probable values of the order parameters have been computed by averaging of the data obtained from one hundred independent numerical experiments at different shear rates $\dot{\gamma}$. As an example, Fig.~\ref{fig_2}a shows the time-dependent order parameter $Q_{6}$ computed for the system at the temperature $T=0.15\,\epsilon/k_{B}$ and at the shear rate of $\dot{\gamma}=0.001\,\tau^{-1}$. Initially, the order parameter $Q_{6}$ takes the small value $\sim0.023$, which corresponds to a disordered system. Further, the glassy system is crystallized after initiation of the shear deformation. Namely, the parameter $Q_{6}$ begins to increase after lag time $\sim70\,\tau$ and at $t>250\,\tau$ the parameter goes to saturation with value $Q_{6}\simeq0.47$, which corresponds to a completely ordered system. The sigmoidal shape of the $Q_{6}(t)$-curve is typical for the activation processes when, for example, the crystallization occurs through the nucleation mechanism~\cite{Mokshin_Galimzyanov_Barrat_2013, Mokshin_Galimzyanov_2013}. A similar scenario of structural ordering is also observed at other shear rates and temperatures with a difference only in the crystallization time scales.

Characteristic directions of the crystalline lattice were initially estimated for the critically-sized nucleus. As we found, there is no alignment of crystal structure of the nuclei along the shear- and/or gradient-directions. Moreover, the correspondence between orientation of the crystal planes and the specific shear directions was also not detected even for the completely ordered systems, where the order parameter $Q_6$ takes its largest value. This is seen from snapshots of the system at $T=0.15\,\epsilon/k_B$ and shear rate $0.001\,\tau^{-1}$ [see Fig.~\ref{fig_2}(b)].

\begin{figure}[ht]
\centering
\includegraphics[width=1.0\linewidth]{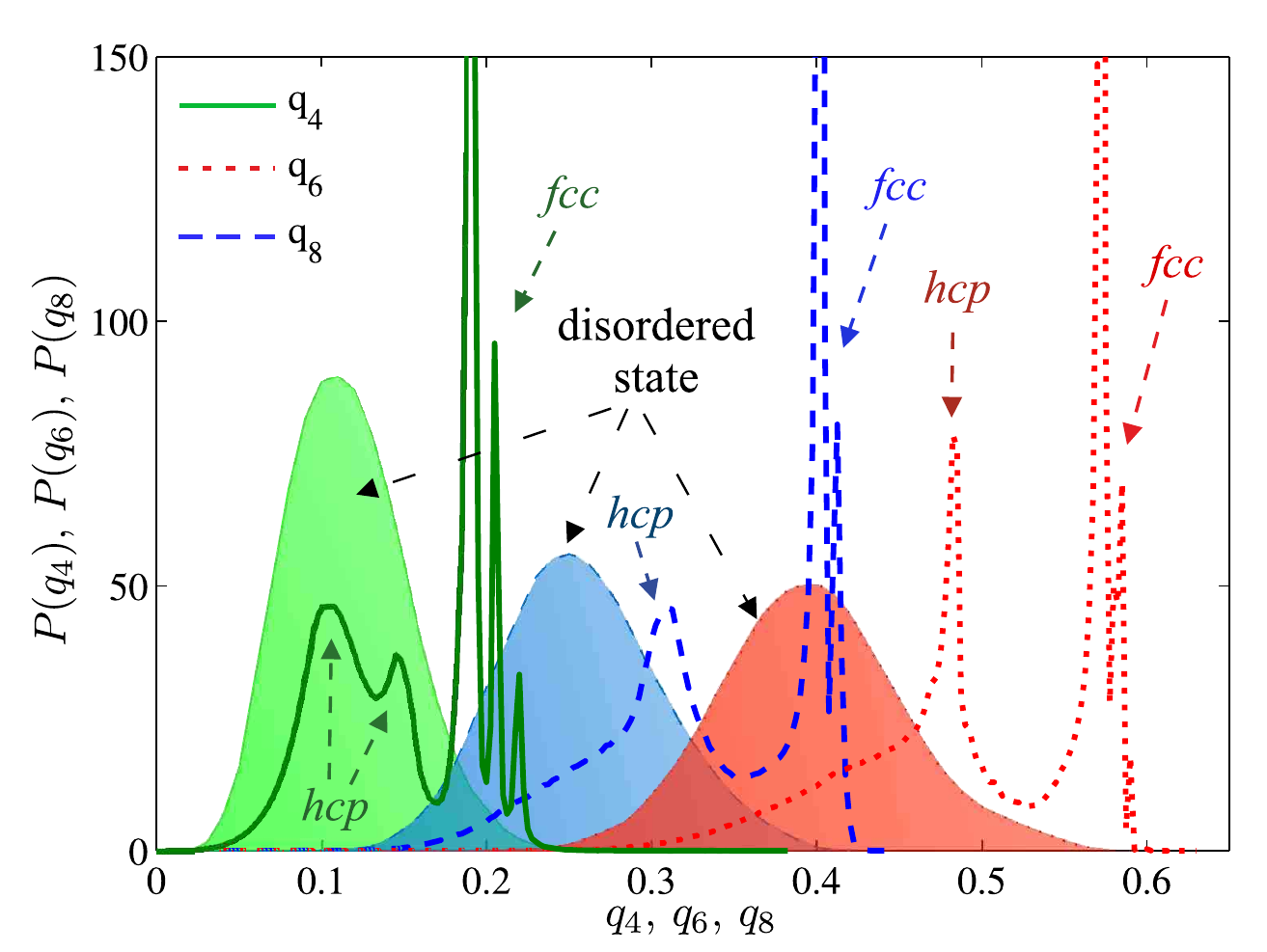}
\caption{(color online) Distributions $P(q_{4})$, $P(q_{6})$, and $P(q_{8})$ evaluated for the Dz-system at the temperature $T=0.15\,\epsilon/k_{B}$ with $Q_{6}\simeq0.47$ crystallized at the shear rate $\dot\gamma=0.001\,\tau^{-1}$. \label{fig_3}}
\end{figure}

Fig.~\ref{fig_3} shows the distributions $P(q_{4})$, $P(q_{6})$, and $P(q_{8})$ for the system at the temperature $T=0.15\,\epsilon/k_{B}$ and at the shear rate $\dot\gamma=0.001\,\tau^{-1}$, when the system is completely crystallized (i.e. when the system corresponds to the state with $Q_{6}\simeq0.47$). The normal distributions $P(q_{4})$, $P(q_{6})$, and $P(q_{8})$ corresponded to the glassy system before initiation of the homogeneous shear are also presented in Fig.~\ref{fig_3} (here, the order parameter is $Q_{6}\simeq0.023$). After initiation of the shear, the structural ordering degree increases. This is also manifested by a change of shapes of $P(q_{4})$, $P(q_{6})$, $P(q_{8})$ and by appearance of peaks located at high values of the parameters $q_{4}$, $q_{6}$, and $q_{8}$. The locations of these peaks in the distributions $P(q_{4})$, $P(q_{6})$, and $P(q_{8})$ correspond to structures with the $fcc$ and $hcp$-symmetries. The fraction of ``$fcc$-particles'' is $\sim60$-$70$\,\%, whereas the fraction of ``$hcp$-particles'' is $\sim30$-$40$\,\%. Moreover, locations of the peaks in the distributions $P(q_{4})$, $P(q_{6})$, and $P(q_{8})$ corresponding to $q_{4}\simeq0.21$, $q_{6}\simeq0.57$, and $q_{8}\simeq0.41$ indicate on the $fcc$-symmetry. The peaks in the distributions at $q_{4}\simeq0.11$, $q_{6}\simeq0.48$, and $q_{8}\simeq0.31$ correspond to the particles forming the $hcp$ lattice~\cite{Mickel_Mecke_2013}. In accordance to with the equilibrium phase diagram of the Dzugutov system, the temperature region along the isobar $p=14\,\epsilon/\sigma^{3}$ includes the crystalline phases with the $fcc$ and $hcp$ lattices~\cite{Dzugutov_1992, Roth_2005}. In particular, at deformation of an amorphous system by shock waves, it was observed in Refs.~\cite{Roth_Denton_2000, Roth_2005} a phase transition into the high-pressure states with $fcc$ or $hcp$ crystalline phase. Both the lattice types appear simultaneously and have a slight difference due to defects in the crystalline structure caused by the shock waves.

\subsection{Critically-sized crystalline nuclei under shear}

\begin{figure*}[t!]
\begin{center}
\includegraphics[width=1.02\linewidth]{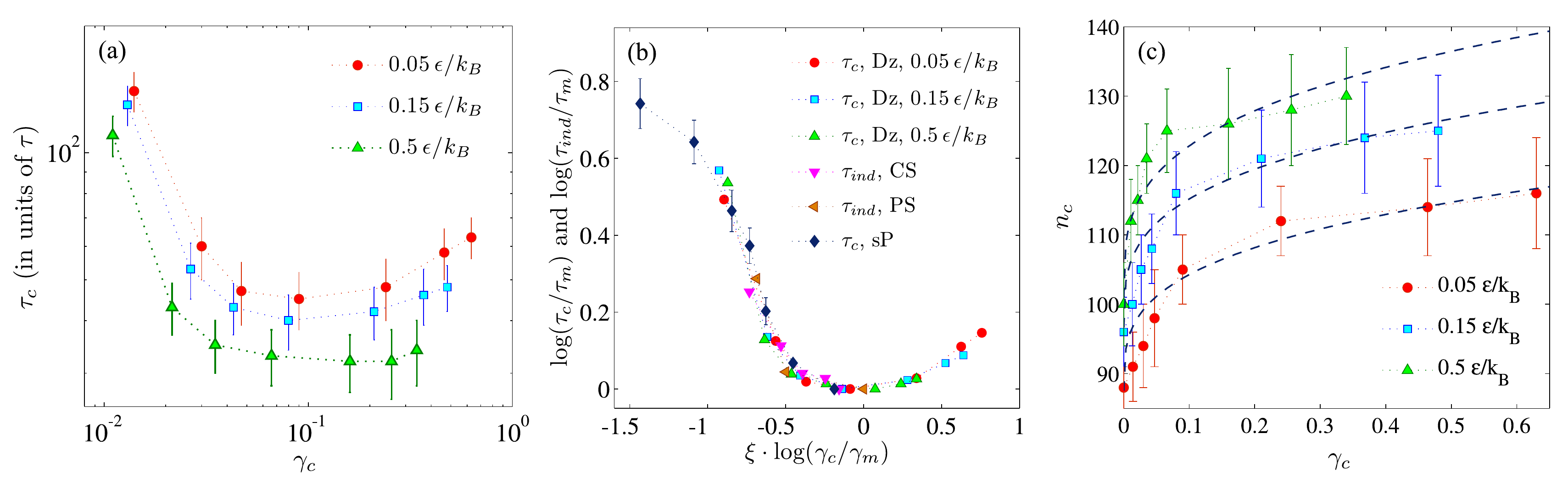}
\caption{(color online) (a) Nucleation time $\tau_c$ as function of critical deformation $\gamma_c$ evaluated for Dz-system at different temperatures. (b) Scaled nucleation time $\tau_c/\tau_m$ and induction time $\tau_{ind}/\tau_m$ as functions of scaled critical deformation $(\gamma_c/\gamma_m)^{\xi}$. Here, $\tau_m$ and $\gamma_m$ are the parameters defined the minimum of the dependence $\tau_c(\gamma_c)$ [and $\tau_{ind}(\gamma_c)$]. Symbols ({\Large$\circ$}), ({\large $\Box$}), and ($\bigtriangleup$) indicate simulation results for the Dzugutov system at different temperatures, whereas symbols ({\footnotesize $\bigtriangledown$}), ({\large $\triangleleft$}), and ({\large $\diamond$}) correspond to experimental data for colloidal system (CS)~\cite{Holmqvist_Dhont_2005} and for two polymer systems: (PS)~\cite{Tribout_Haudin_1996} and (sP)~\cite{Coccorullo_Titomanlio_2008}. (c) Critical size $n_c$ vs. $\gamma_c$. Dashed curves are results of fit by Eq.~(\ref{eq_nc3}). \label{fig_4}}
\end{center}
\end{figure*}

Results of the cluster analysis reveal that the structural ordering in the glassy Dz-system occurs through the formation of small-sized crystalline nuclei consisting mainly of $30$-$50$ particles. By the mean-first-passage-time method, the critical size $n_c$ and the nucleation time $\tau_c$ have been evaluated for the system at different temperatures and shear rates~\cite{Mokshin_Galimzyanov_2012, Mokshin_Galimzyanov_Barrat_2013}. Fig.~\ref{fig_4}(a) shows the nucleation time $\tau_c$ as function of the critical deformation $\gamma_c$ evaluated at different temperatures. We found different regimes in $\gamma_c$-dependencies of $\tau_c$. Namely, the nucleation time decreases at small and moderate shear rates and it increases at high shear rates. The nucleation time decreases from $\tau_{c}\simeq125\pm15\,\tau$ to $\tau_{c}\simeq40\pm6\,\tau$ with increasing shear rate from $\dot{\gamma}=0.0001\,\tau^{-1}$ ($\gamma_{c}\simeq0.0125$) to $\dot{\gamma}=0.002\,\tau^{-1}$ ($\gamma_{c}\simeq0.08$). The nucleation time increases from $\tau_{c}\simeq41\pm5\,\tau$ to $\tau_{c}\simeq53\pm10\,\tau$ at increasing shear rate from $\dot{\gamma}=0.005\,\tau^{-1}$ ($\gamma_{c}\simeq0.205$) to $\dot{\gamma}=0.01\,\tau^{-1}$ ($\gamma_{c}\simeq0.53$) in the considered temperature range $T\in[0.05;\,0.5]\,\epsilon/k_{B}$. Such non-monotonic $\gamma_c$-dependence of the nucleation time $\tau_c$ is due to antagonistic impact of shear flow on the nucleation process. This effect was also discussed in Refs.~\cite{Mokshin_Barrat_2010,Mokshin_Galimzyanov_Barrat_2013}. As discussed in Ref.~\cite{Mokshin_Galimzyanov_Barrat_2013}, such non-monotonic behavior of $\tau_c$ is due to impact of shear deformation on the kinetic and thermodynamic aspects of the crystal nucleation in this system. Namely, a slow shear-flow accelerates the nucleation through the attachment rate, whereas the high shear rates destabilize the critical nuclei and reduce the probability of the particle attachment. On the other hand, a shear flow gives rise to ``pressure anisotropy''~\cite{Reguera_Rubi_2003} manifested in the fact that the diagonal components of the pressure tensor begin to differ. This effect leads to anisotropy of the interfacial free energy. Moreover, the shear deformation introduces an elastic energy into the system and, thereby, it has an impact on the thermodynamic characteristics of crystal nucleation (the nucleation barrier, the interfacial free energy, etc.)~\cite{Mura_Zaccone_2016,Lifshitz_Pitaevskii_2006}.

Although there are known experimental data characterizing the crystal nucleation in amorphous systems under shear, comparison of different experimental data and results of numerical simulation is a non-trivial task. In fact, these systems are characterized by various physical, chemical and rheological properties, and the systems are studied at different thermodynamic and deformation conditions. Nevertheless, there are common regularities in structural ordering in these systems. Namely, the slow shear deformations accelerate the ordering, whereas the large shear rates suppress the crystallization. As a result, the strain-dependencies of the crystallization (nucleation) time must be characterized by a minimum. Therefore, it is reasonable to propose that such the dependencies can be represented in the unified scaled form:
\begin{equation} \label{eq_univdep}
\frac{\tau_{c}}{\tau_{m}}=f\left[\left(\frac{\gamma_{c}}{\gamma_{m}}\right)^{\xi}\right]
\end{equation}
Here, $f[\ldots]$ is a universal function, $\tau_m$ is the location of the minimum in the dependence $\tau_c(\gamma_c)$, whereas $\gamma_m$ is the value of the critical deformation $\gamma_c$ at the time $\tau_m$. Further, the dimensionless positive parameter $\xi$ can be considered as the characteristic of glass-forming ability of the system~\cite{Mokshin_Galimzyanov_2015,Weingartner_Nussinov_2017,Mokshin_Galimzyanov_2017}.

In Fig.~\ref{fig_4}(b), the scaled time $\tau_c/\tau_m$ is presented as function of the scaled critical deformation $(\gamma_c/\gamma_m)^{\xi}$ for different systems under shear: our simulation results for the Dzugutov system (Dz); experimental data for colloidal system (CS)~\cite{Holmqvist_Dhont_2005} and two polymer systems (PS)~\cite{Tribout_Haudin_1996} and (sP)~\cite{Coccorullo_Titomanlio_2008}. For the case of CS and PS systems, the data for the induction time $\tau_{ind}$ vs. the critical deformation $\gamma_c$ were taken. As seen from Fig.~\ref{fig_4}(b), all the data collapse onto unified curve. This is evidence that the scaling relation (\ref{eq_univdep}) is valid. Here, the parameter $\xi$ takes the next values: $\xi=0.935$ for the Dzugutov system, $1.282$ for the CS-system, $2.33$ for the PS-system and $1.89$ for the sP-system. Thus, the parameter takes the lowest value for the atomistic single-component Dzugutov system, and it takes the larger values for the polymer and colloidal systems with better glass-forming ability properties.

It should be noted that the crystal nucleation in amorphous solid (glass) differs from crystal nucleation in supercooled liquid. At low and moderate levels of supercooling corresponding to a liquid (supercooled) phase, the nucleation rate is determined mainly by the thermodynamic factor~\cite{Mokshin_Galimzyanov_Barrat_2013,Mokshin_Galimzyanov_2015,Kashchiev_Nucleation_2000}. For a glassy phase with deep levels of supercooling, the nucleation is driven by the kinetic features of the system. Competition between the thermodynamic and kinetic aspects is clearly manifested in appearance of the characteristic maximum in the temperature dependence of the nucleation rate~\cite{Kashchiev_Nucleation_2000,Kelton_Greer_2010,Kalikmanov_2012}. Influence of the shear on crystallization of the supercooled liquid and glass is also different. In particular, for the case of a liquid, the shear deformation increases the viscosity and, thereby, nucleation is suppressed~\cite{Mura_Zaccone_2016,Zaccone_Morbidelli_2009}. For a glassy system, the shear increases \emph{effectively} the mobility of particles and, thereby, the shear decreases the viscosity.

The system evolves at temperatures much below the glass transition temperature $T_g$. And, therefore, detection of the nucleation event for the system at zeroth shear may seem surprising. Actually, features of the microscopic kinetics of a glass changes with moving over phase diagram for the range of high pressures. Namely, at high pressures the structural relaxation as well as the transition of glassy system into a state with the lower free energy proceeds over shorter time scales~\cite{Shen_Reed_2016}. As we found, the nucleation event are detectable within simulation time scales for the Dz-system at the thermodynamic states with pressures $p\geq12\,\epsilon/\sigma^3$~\cite{Mokshin_Galimzyanov_2015}.

In Fig.~\ref{fig_4}(c), the quantity $n_{c}$ as function of the critical deformation $\gamma_{c}$ evaluated at different temperatures is presented. At the absence of shear, for the considered temperature range $T\in[0.05;\,0.5]\,\epsilon/k_{B}$, the critical size increases with temperature by several tens of particles. This is in agreement with classical nucleation theory~\cite{Kashchiev_Nucleation_2000,Kelton_Greer_2010}, according to which $n_c=(32\pi\gamma_{\infty}^3)/(3\rho_{c}|\Delta\mu|^3)$. Here, $\gamma_{\infty}$ is the interfacial free energy, $|\Delta\mu|$ is the difference of the chemical potential per phase unite in the melt (glass) and the crystal. Then, taking into account that the chemical potential difference $|\Delta\mu|$ is proportional the supercooling, $\Delta T=T_m-T$:
\begin{equation} \nonumber
|\Delta\mu| \sim\frac{\Delta T}{T_m},
\end{equation}
one has that the critical size $n_c$ increases with increasing temperature. Further, with increase of shear rate within the range $\dot\gamma\in[0.0001;\,0.01]\,\tau^{-1}$, the critical size increases from $n_{c}\simeq88$ to $n_{c}\simeq130$ particles. As seen from Fig.~\ref{fig_4}(c), at the low shear rates $\dot{\gamma}<0.001\,\tau^{-1}$, the nuclei appear at critical deformation $\gamma_{c}=0.02$, whereas at high shear rates formation of the nuclei occurs at the larger deformations. Namely, at the shear rate $\dot{\gamma}=0.01\,\tau^{-1}$, the nuclei appear at the critical deformation $\gamma_{c}>0.3$. The large values of $\gamma_{c}$ are due to that the high shear rates inhibit the crystal nucleation, that was previously discussed in Refs.~\cite{Mokshin_Galimzyanov_Barrat_2013,Mura_Zaccone_2016}.

\begin{figure}[t!]
\begin{center}
\includegraphics[width=0.85\linewidth]{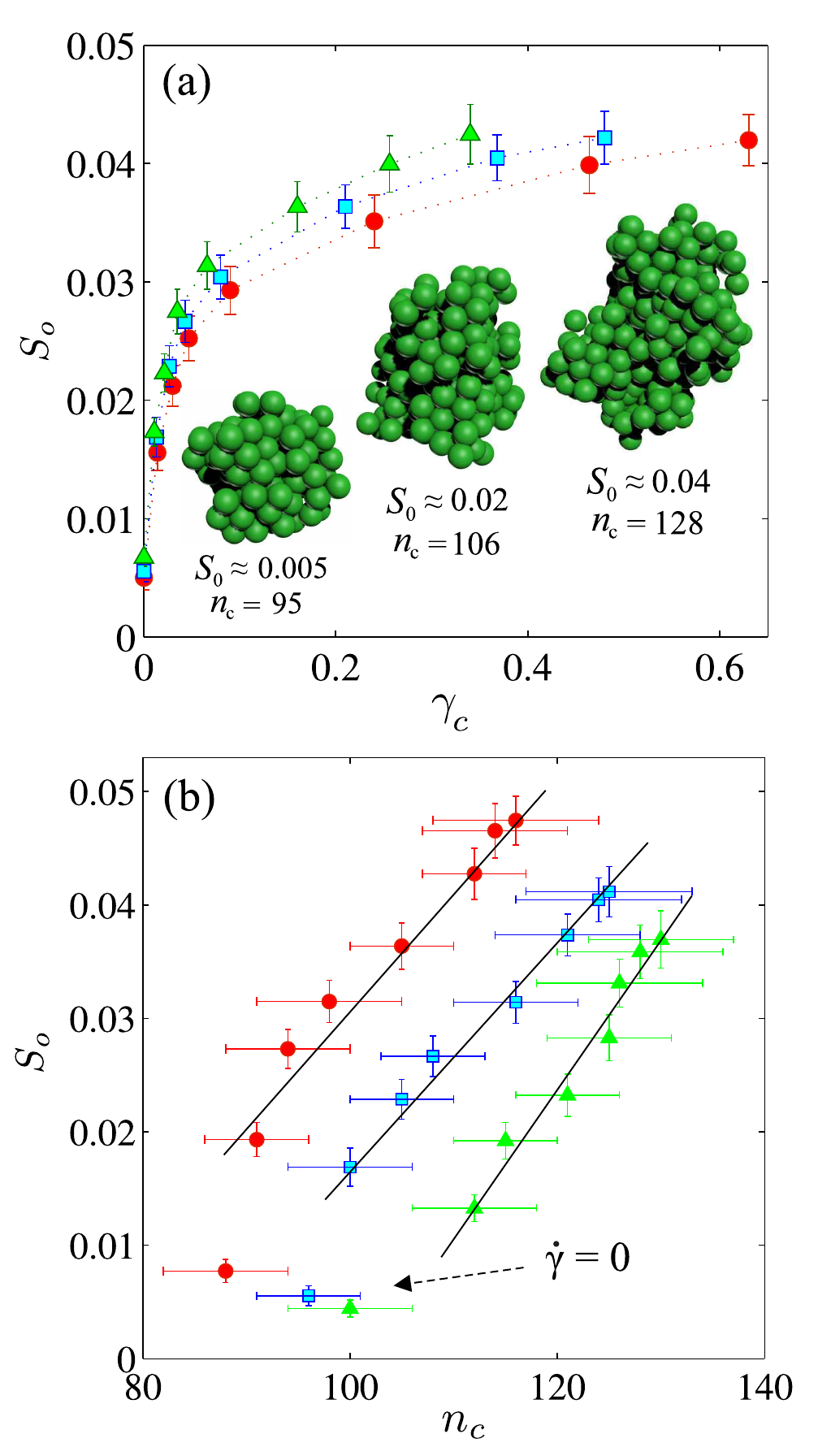}
\caption{(color online) (a) Asphericity parameter $S_{0}$ as function of the critical deformation $\gamma_{c}$. Insets show the critically-sized nuclei with various sizes and shapes. (b) Correlation between the asphericity parameter $S_{0}$ and the critical size $n_{c}$ at different temperature of the system. Here, circles correspond to the system at $T=0.05\,\epsilon/k_{B}$, squares correspond to $T=0.15\,\epsilon/k_{B}$, and triangles correspond to $T=0.5\,\epsilon/k_{B}$. \label{fig_5}}
\end{center}
\end{figure}

For the crystalline nuclei with sizes $\leq100$ particles, it is quite reasonable within the statistical treatment to characterize their shape as a quantity averaged over data of independent experiments. This averaging procedure is denoted by $\left\langle...\right\rangle$ in Eq.~(\ref{eq_So_parameter}). As seen from Fig.~\ref{fig_5}(a), for a shear-free system as well as for the system at small shear rates $\dot{\gamma}<0.001\,\tau^{-1}$, the asphericity parameter takes the values $S_{0}<0.02$. It indicates that the critically-sized nucleus approximate a symmetric shape close to be a spherical. Further, the quantity $S_{0}$ increases from $\simeq0.005$ to $S_{0}\simeq0.045$ with the increasing critical deformation $\gamma_{c}$, that is due to increase of the nucleus shape asphericity. Homogeneous steady-state shear promotes local rearrangements of the particles and leads effectively to an increase of particle mobility. Therefore, the system temperature takes a sense of the effective parameter~\cite{Cugliandolo_Kurchan_1997}, which increases with increase of the shear rate (Fig. 4 in Ref.~\cite{Mokshin_Galimzyanov_Barrat_2013}). Moreover, there is anisotropy of interfacial free energy due to shear. Then, the system with a specific characteristic interparticle interaction must be characterized by a limit value of curvature of the solid (crystal)-liquid interface; while the geometries with larger values of curvature will be unstable. As a result, a plateau in the $\gamma_c$-dependence of the asphericity parameter appears and saturation of the $\gamma_c$-dependence of the critical size is observed at high shear rates. Therefore, it is remarkable that the $\gamma_{c}$-dependence of $S_{0}$ is correlated with such the dependence of $n_{c}$. One can see in Fig.~\ref{fig_5}(b) that both the quantities $S_{0}$ and $n_{c}$ are correlated, and a larger value of the asphericity parameter $S_{0}$ corresponds to a larger value of the critical size $n_{c}$. This scenario differs from crystal nucleation without external mechanical drive, where $S_{0}\propto1/n_{c}$~\cite{Reinhardt_Doye_2012}.

Projections of the pair-distribution function $g(r)$ onto the shear-gradient $xy$ and the shear-vorticity $xz$ planes were defined for particles of the critically-sized nucleus as the next:
\begin{equation}
g_{c}(x,y)=\left\langle\frac{1}{n_{c}}\sum_{i=1}^{n_{c}}\sum_{j>i}^{n_{c}}\delta\left(\vec{r}-\vec{r}_{ij}(x,y)\right)\right\rangle\label{eq_Gxy}
\end{equation}
and
\begin{equation}
g_{c}(x,z)=\left\langle\frac{1}{n_{c}}\sum_{i=1}^{n_{c}}\sum_{j>i}^{n_{c}}\delta\left(\vec{r}-\vec{r}_{ij}(x,z)\right)\right\rangle.\label{eq_Gxz}
\end{equation}
Here, $\langle...\rangle$ means an averaging over data of independent simulation runs. At evaluation of the distributions $g_{c}(x,y)$ and $g_{c}(x,z)$ the positions of particles were determined with respect to geometric center of a nucleus.

\begin{figure*}[t!]
\begin{center}
\includegraphics[width=1.0\linewidth]{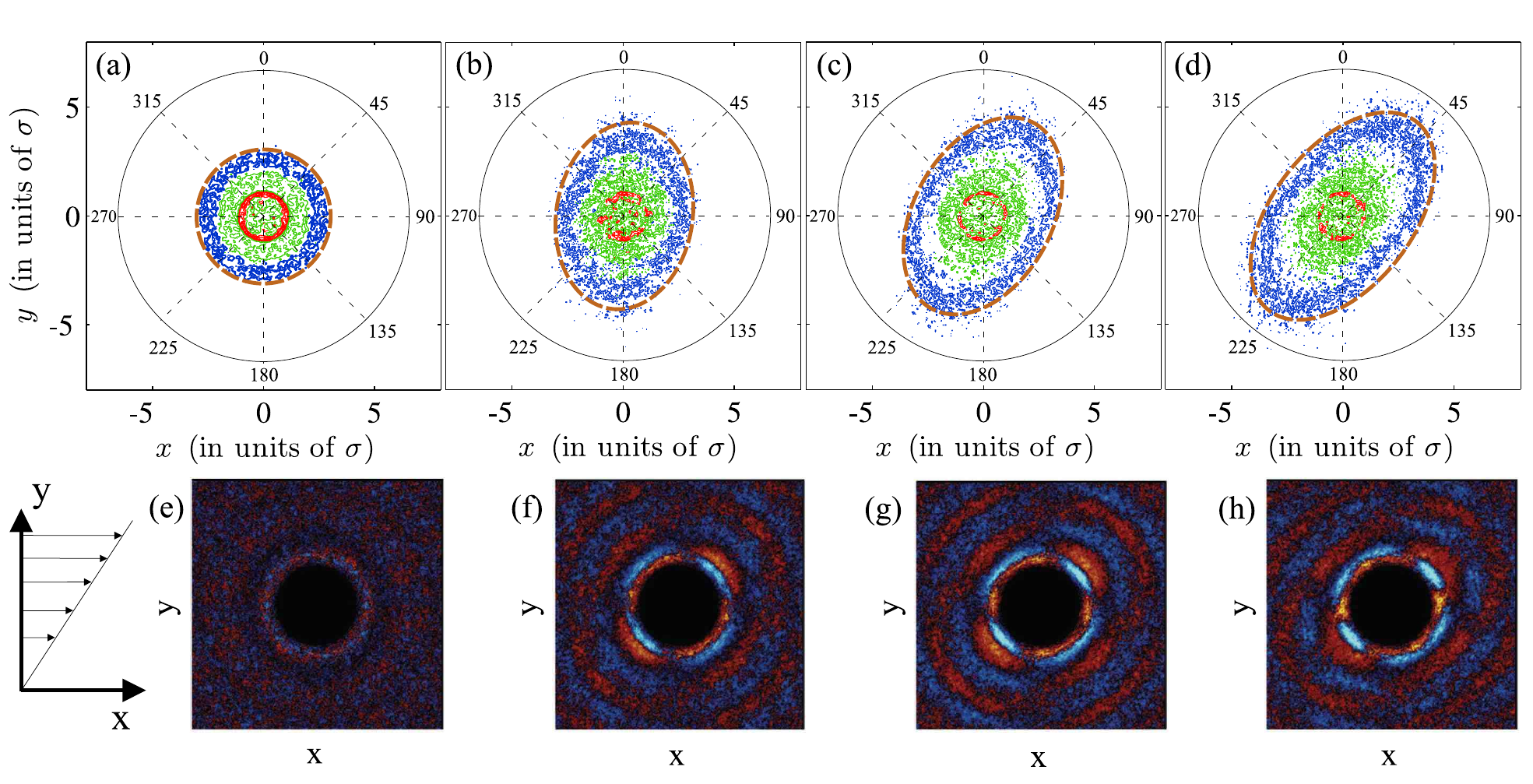}
\caption{(color online) \textbf{Top}: projections of the pair distribution function $g_{c}(x,y)$ for the particles of a critically-sized nucleus received for the system at the temperature
$T=0.15\,\epsilon/k_{B}$ and at the shear rates: (a) zeroth shear rate,
$\dot{\gamma}=0$; (b) $\dot{\gamma}=0.0005\,\tau^{-1}$ (critical deformation $\gamma_{c}\sim3$\%); (c)
$\dot{\gamma}=0.005\,\tau^{-1}$ ($\gamma_{c}\sim20$\%); (d)
$\dot{\gamma}=0.01\,\tau^{-1}$ ($\gamma_{c}\sim50$\%). External boundaries of the distribution $g_{c}(x,y)$ are shown by the dashed circle and ellipses. \textbf{Bottom}: projections of the pair distribution function onto shear-gradient $xy$-plane obtained for hard-sphere glassy systems at different shear strains: (e) $1$\%, (f) $10$\%, (g) $20$\% and (h) $60$\% (taken from Refs.~\cite{Koumakis_Petekidis_2012,Koumakis_Petekidis_2016}).} \label{fig_6}
\end{center}
\end{figure*}

\begin{figure}[t!]
\begin{center}
\includegraphics[width=1.0\linewidth]{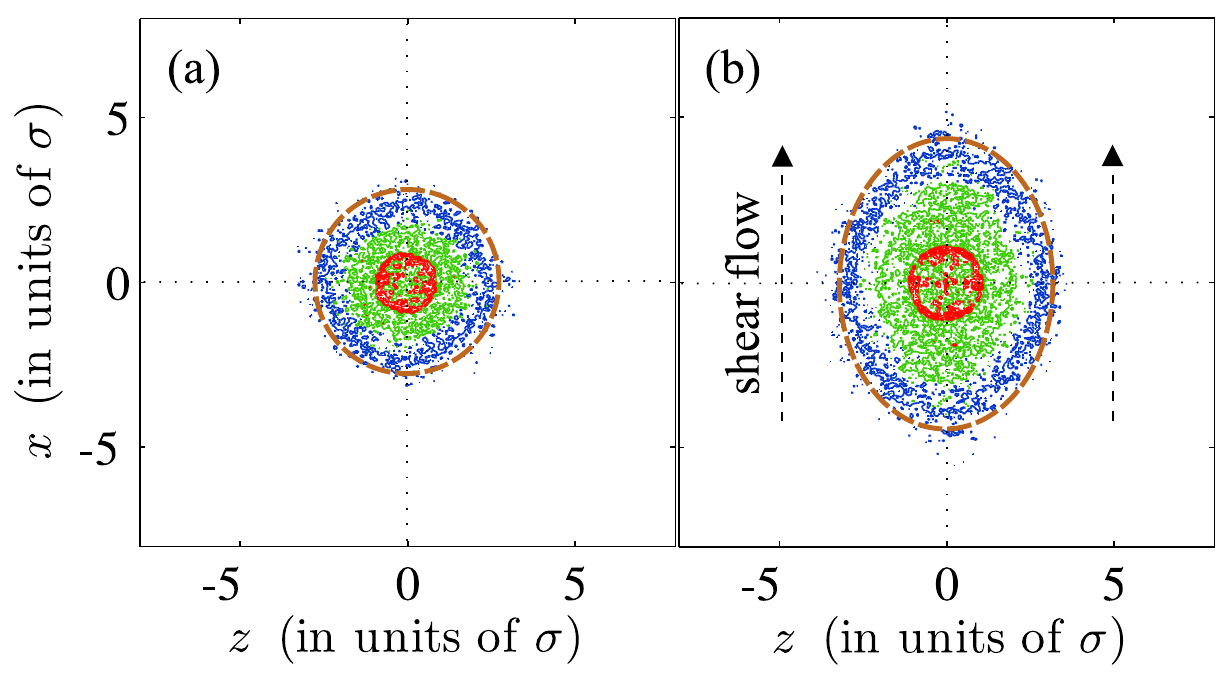}
\caption{(color online) Projections of the pair distribution function $g_{c}(x,z)$ onto the $xz$-planes
for the particles of a critically-sized nucleus received for the system at the temperature
$T=0.15\,\epsilon/k_{B}$ and at the shear rates: (a)
$\dot{\gamma}=0$; (b) $\dot{\gamma}=0.01\,\tau^{-1}$. External boundaries of the distribution $g_{c}(x,z)$ are shown by the dashed circle and ellipse.}
\label{fig_7}
\end{center}
\end{figure}

The distributions $g_{c}(x,y)$ for the system at the temperature $T=0.15\,\epsilon/k_{B}$ and at the shear rates $\dot\gamma=0$, $0.0005$, $0.005$, and $0.01\,\tau^{-1}$ are shown in Fig.~\ref{fig_6}. Further, Fig.~\ref{fig_7} presents the distributions $g_{c}(x,z)$ for the shear-free system at the temperature $T=0.15\,\epsilon/k_{B}$ and for the system at the shear rate $0.01\,\tau^{-1}$. As seen, the closed lines corresponding to the first, second and third coordinations are well recognizable in these distributions. The lines are presented by red, green and blue colors, respectively. At the absence of a shear, the contours of the distributions $g_{c}(x,y)$ and $g_{c}(x,z)$ are circular, that is evidence of a spherical shape of the nucleus. As can be seen from contours of $g_{c}(x,y)$ and $g_{c}(x,z)$, the nucleus asphericity becomes more pronounced with shear increase [Figs.~\ref{fig_6}(b), \ref{fig_6}(c), \ref{fig_6}(d), and Fig.~\ref{fig_7}(b)]. This is also observed at other considered temperatures, not presented in Figs.~\ref{fig_6} and~\ref{fig_7}.

The contours of the distributions $g_{c}(x,y)$ and $g_{c}(x,z)$ change forms from circular to elliptical at increasing shear rate $\dot\gamma$. The long axis of the elliptic contour of $g_{c}(x,y)$ is oriented with respect to the gradient direction [see Figs.~\ref{fig_6}(b), \ref{fig_6}(c) and \ref{fig_6}(d)]. At the same time, the orientation of this ellipse within the $xz$-plane is not observed, that is seen, in particular, in Fig.~\ref{fig_7}. Namely, the long axis of the ellipse in the distribution $g_{c}(x,z)$ increases with increasing shear rate only in the shear direction [see Fig.~\ref{fig_7}(b)], whereas the small axis of the ellipse is practically unchanged both in the $xy$ and $xz$-planes. This is direct evidence that at high shear rates the nuclei are characterized by an elongated ellipsoidal shape. Note that this is in agreement with results of Refs.~\cite{Blaak_PRL_93_2004, Somani_2005, Olmsted_PRL_103_2009, Wang_Liu_2016}.
\begin{figure}[ht]
\centering
\includegraphics[width=1.0\linewidth]{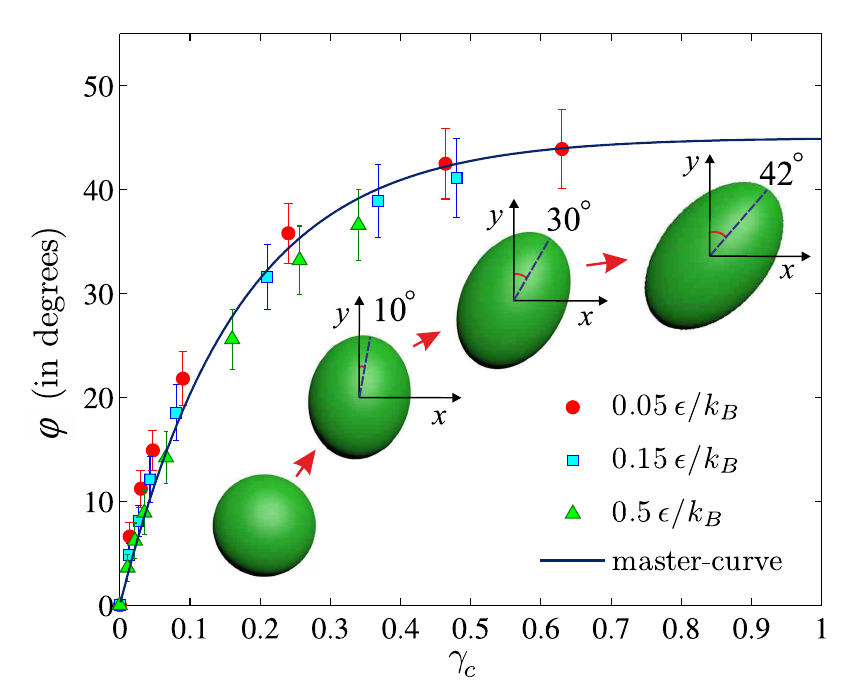}
\caption{(color online) Tilt angle $\varphi$ of a critically-sized nucleus as a function of the critical deformation $\gamma_{c}$ at different
temperatures. The solid curve is fit by Eq.~(\ref{eq_tilt_angle}). Insets: schematic illustrations showing the change of orientation and shape of the nuclei at various $\gamma_{c}$. \label{fig_8}}
\end{figure}

The tilt angle $\varphi$ between the gradient direction and the longest axis of the (ellipsoid-like) nucleus was determined directly from contours of the distribution $g_{c}(x,y)$. In Fig.~\ref{fig_8}, the $\gamma_{c}$-dependencies of $\varphi$ obtained at different shear rates and temperatures are shown. The tilt angle $\varphi$ increases with the increasing critical deformation $\gamma_{c}$ according to the power-law dependence
\begin{equation}
\varphi(\gamma_{c})=\varphi_{0}\left\{1-\exp\left(-\mathcal{D}\gamma_{c}\right)\right\}.\label{eq_tilt_angle}
\end{equation}
Here, $\mathcal{D}$ and $\varphi_{0}$ are the positive fitting parameters. As it appears, both the parameters are independent on the temperature $T$ for the considered temperature range and take the values $\mathcal{D}=6.0\pm0.05$ and $\varphi_{0}=45^{\circ}\pm3^{\circ}$, respectively. The function $\varphi(\gamma_{c})$ goes to saturation faster at the higher value of the parameter $\mathcal {D}$. The parameter $\varphi_{0}$ is the limit value of $\varphi$. Namely, at the large deformations (in our case at $\gamma_{c}>0.6$) the long axis of the elliptic contour in the distribution $g_{c}(x,y)$ is oriented at the angle $\varphi_{0}\approx45^{\circ}$ with respect to the gradient direction (see also Refs.~\cite{Nosenko_Ivlev_2012, Dasgupta_Procaccia_2012, Koumakis_Petekidis_2016}). As it follows from Refs.~\cite{Koumakis_Petekidis_2012,Koumakis_Petekidis_2016}, a pronounced anisotropy in microscopic structure of hard-sphere suspensions and glasses is observed with increasing shear strain. It was found in Refs.~\cite{Koumakis_Petekidis_2012,Koumakis_Petekidis_2016} that with increasing shear deformation the particles density starts to be dependent on direction, and the so-called stretching and compression directions appear. It is remarkable that the contours of this distribution [see Fig.~\ref{fig_6}(e),~\ref{fig_6}(f),~\ref{fig_6}(g) and~\ref{fig_6}(h)] are similar to the contours of the distribution $g_{c}(x,y)$ [see Fig.~\ref{fig_6}(a),~\ref{fig_6}(b),~\ref{fig_6}(c) and~\ref{fig_6}(d)], where the angle between the stretching direction and the gradient direction goes to the limit value $\approx45^{\circ}$. On the other hand, it was experimentally found by Nosenko \emph{et al.}~\cite{Nosenko_Ivlev_2012} for strongly coupled liquid that the microscopic structure of the system at influence of shear becomes anisotropic; and compressive and stretching axes tend to be oriented at the angle of $\pm45^{\circ}$ respective to the direction of flow. Evidently, such ellipticity of the distribution $g_{c}(x,y)$ arising at steady shear has an impact on the crystal nucleation processes and is also responsible for the critically-sized nuclei asphericity.

\subsection{Size and Shape of Critically-Sized Nuclei}

\begin{figure}[ht]
\centering
\includegraphics[width=1.0\linewidth]{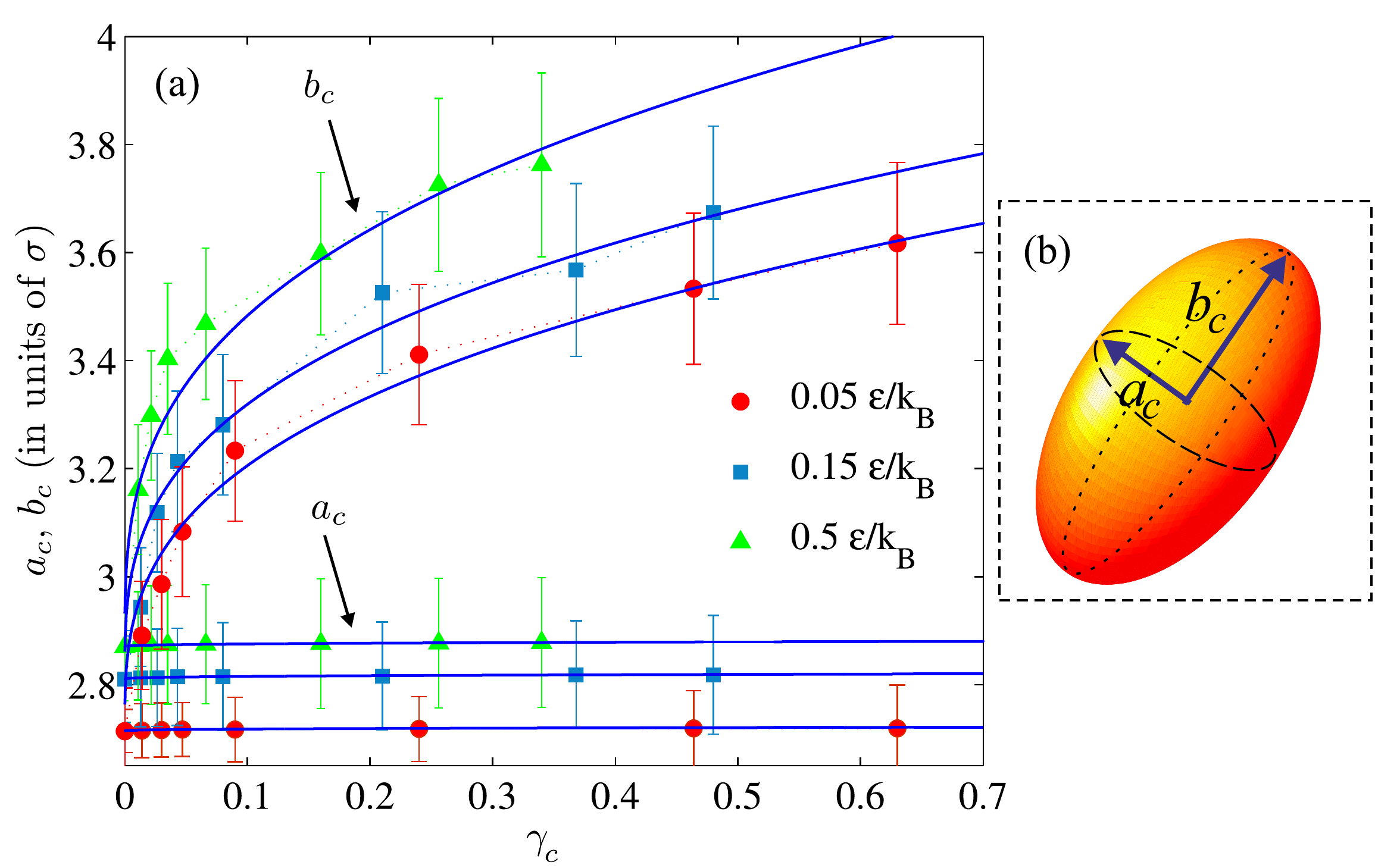}
\caption{(color online) (a) Quantities $a_{c}$ and $b_{c}$ as functions of the critical deformation $\gamma_{c}$. The solid curves are fit by Eqs.~(\ref{eq_fit_ac}) and (\ref{eq_fit_bc}). (b) Schematic image of the ellipsoidal shape, where $a_{c}$ and $b_{c}$ are small and large semiaxes determined from the distribution $g_{c}(x,y)$.}
\label{fig_9}
\end{figure}

Let $a_{c}$ and $b_{c}$ be the small and large semiaxes of the elliptic contours in the distribution $g_{c}(x,y)$, respectively. The quantities $a_{c}$ and $b_{c}$ are evaluated from the distribution $g_{c}(x,y)$ at different critical deformation $\gamma_{c}$. As seen from Fig.~\ref{fig_9}, the small semiaxis $a_{c}$ is practically unchanged with the increasing critical deformation $\gamma_{c}$, while the quantity $b_{c}$ increases with $\gamma_{c}$. Remarkably, the $\gamma_{c}$-dependencies of the quantities are well reproduced by
\begin{equation}
a_{c}(\gamma_{c})=a_{0}\left\{1+\mathcal{A}\gamma_{c}^{\kappa}\right\}\label{eq_fit_ac}
\end{equation}
and
\begin{equation}
b_{c}(\gamma_{c})=a_{0}\left\{1+\mathcal{B}\gamma_{c}^{\kappa}\right\}.\label{eq_fit_bc}
\end{equation}
Here, $a_{0}$ is the radius of a spherical nucleus at $\dot\gamma=0$, and it is $a_{0}\simeq2.71$, $2.81$, and $2.87\,\sigma$ for temperatures $T=0.05$, $0.15$, and $0.5\,\epsilon/k_{B}$, respectively. The dimensionless parameters $\mathcal{A}$ and $\mathcal{B}$ take the fixed positive values $\mathcal{A}\simeq0.004\pm0.001$ and $\mathcal{B}\simeq0.42\pm0.04$ for the considered temperature range $T\in[0.05;\,0.5]\,\epsilon/k_{B}$. The dimensionless parameter $\kappa$ is defined by the thermodynamic properties of a system and takes positive value $\kappa=1/3$ for considered temperatures. We found that the ratio $\mathcal{B}/\mathcal{A}$ is $\approx105$ and, consequently, the asphericity of a nucleus shape increases mainly due to increase of the large semiaxis $b_{c}$.

From Eqs.~(\ref{eq_fit_ac}) and (\ref{eq_fit_bc}) one obtains the $\gamma_{c}$-dependence of the critical size $n_{c}$:
\begin{eqnarray} \label{eq_nc2}
n_{c}(\gamma_{c})&=&\frac{4}{3}\pi\rho_{c}\left[a_{c}(\gamma_{c})\right]^{2}\left[b_{c}(\gamma_{c})\right] \\
&=& \frac{4}{3}\pi\rho_{c}a_{0}^{2}b_{0}\{1+
\mathcal{B}\gamma_{c}^{\kappa}+
2\mathcal{A}\gamma_{c}^{\kappa}
\nonumber \\ &+& 2\mathcal{A}\mathcal{B}\gamma_{c}^{2\kappa}+
\mathcal{A}^{2}\gamma_{c}^{2\kappa}+
\mathcal{A}^{2}\mathcal{B}\gamma_{c}^{2\kappa}\}. \nonumber
\end{eqnarray}
Since the parameter $\mathcal{A}$ takes small values in the case of the Dz-system ($\mathcal{A}\simeq0.004$) and taking into account $b_{c}\simeq a_{c}$ at $\dot\gamma=0$, Eq.~(\ref{eq_nc2}) can be rewritten in the following simplified form:
\begin{equation}
n_{c}(\gamma_{c})=n_{c}(\dot\gamma=0)\left\{1+\mathcal{B}\gamma_{c}^{\kappa}\right\},\label{eq_nc3}
\end{equation}
where
\begin{equation}
n_{c}(\dot\gamma=0)=\frac{4}{3}\pi\rho_{c}a_{0}^{3}.\label{eq_nc4}
\end{equation}
Here, $\rho_{c}\simeq1.04\pm0.02\,\sigma^{-3}$ is the numerical density of the crystalline phase. Fig.~\ref{fig_4}(c) shows the fit of the simulation results by Eq.~(\ref{eq_nc3}) at given values of $\mathcal{B}$, $\kappa$, and $a_{0}$. It can be seen from Fig.~\ref{fig_4}(c), Eq.~(\ref{eq_nc3}) reproduces correctly the $\gamma_{c}$-dependence of the critical size $n_{c}$ at the considered temperatures.

\begin{figure*}[ht]
\centering
\includegraphics[width=1.0\linewidth]{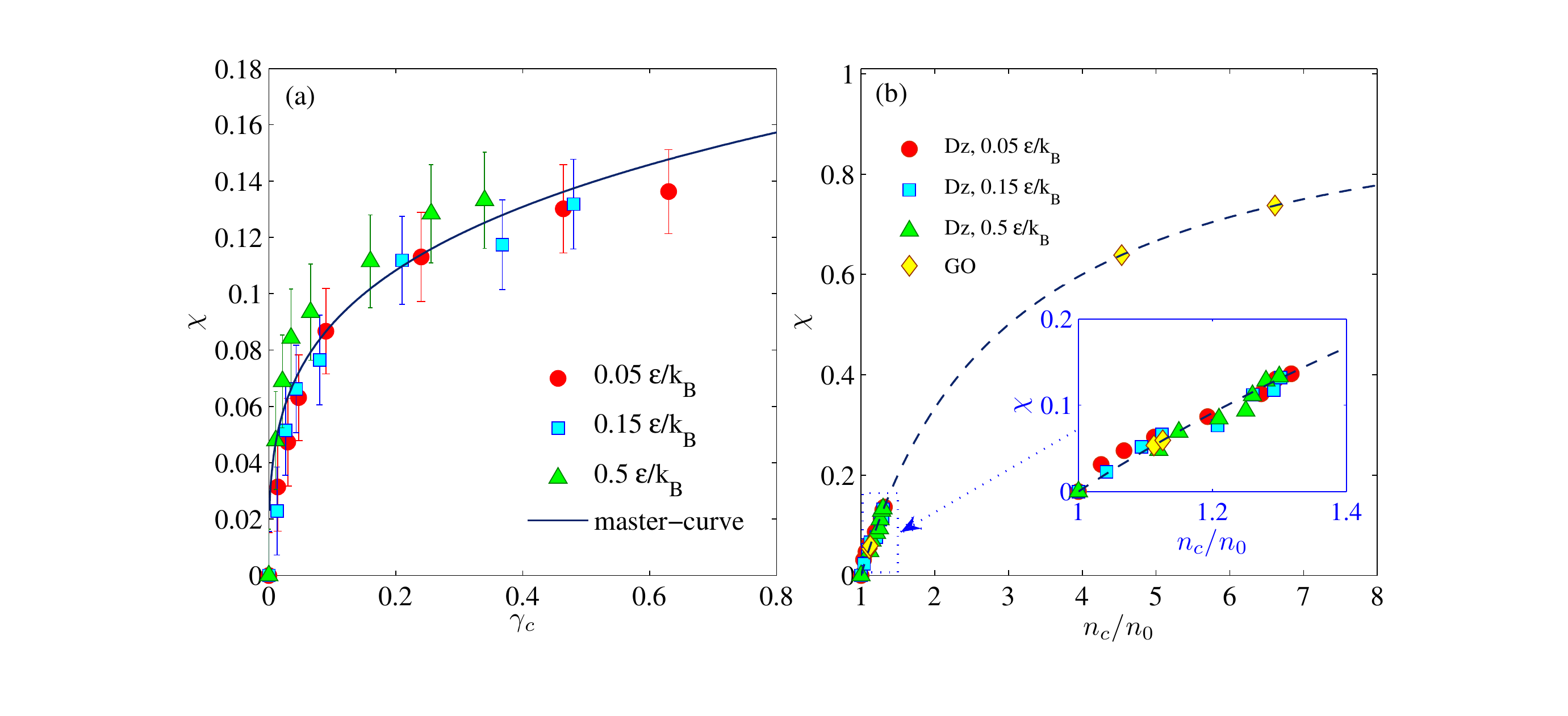}
\caption{(color online) (a) Quantity $\chi$ as function of the critical deformation $\gamma_{c}$. The solid curve is fit by Eq.~(\ref{eq_chi_tau}). (b) Quantity $\chi$ as function of the scaled critical size $n_{c}/n_{0}$, where $n_{0}\equiv n_{c}(\dot\gamma=0)$ is the critical size at zeroth shear rate. The dashed curve is fit by Eq.~(\ref{eq_chi_nc_fit}). The data for elongated critically-sized crystalline nuclei marked as $GO$ are taken from Refs.~\cite{Olmsted_PRL_103_2009,Graham_Olmsted_2010}.}
\label{fig_10}
\end{figure*}

For the quantitative characterization of the nuclei shape we define the deformation parameter
\begin{equation}
\chi=\frac{b_{c}-a_{c}}{b_{c}+a_{c}}.\label{eq_aspR}
\end{equation}
For the case $b_{c}\geq a_{c}$ the parameter $\chi$ takes the values from the range $[0;\,1]$. Namely, if a nucleus is characterized by a perfect spherical shape, we have $\chi=0$, whereas for an elongated nucleus one has $\chi\rightarrow1$.

The $\gamma_{c}$-dependencies of the parameter $\chi$ at different temperatures are shown in Fig.~\ref{fig_10}(a). An increase of $\chi$ with the critical deformation $\gamma_{c}$ is mainly due to increase of the value $b_{c}$. It should be noted that the $\gamma_{c}$-dependencies of $S_{0}$ and $\chi$ are correlated [see Figs.~\ref{fig_4}(c) and~\ref{fig_10}(a)]. Further, the results reveal that the $\gamma_{c}$-dependencies of the parameter $\chi$ are reproduced by
\begin{equation}
\chi(\gamma_{c})=\frac{\mathcal{B}\gamma_{c}^{\kappa}}{2+\mathcal{B}\gamma_{c}^{\kappa}},\label{eq_chi_tau}
\end{equation}
which follows from Eqs.~(\ref{eq_fit_ac}), (\ref{eq_fit_bc}) and (\ref{eq_aspR}). Here, $\mathcal{B}\simeq0.42\pm0.04$ and $\kappa\simeq1/3$ for the considered temperatures. We found that the $\gamma_{c}$-dependencies of the parameter $\chi$ goes to saturation at high critical deformations, and  extremely high shear rates do not lead to the high asphericity of a nucleus.

Moreover, Fig.~\ref{fig_10}(b) shows the parameter $\chi$ as function of the scaled critical size $n_{c}/n_{0}$. Here, the quantity $n_{0}$ is the critical size, which is evaluated for the system at absence of a shear, $n_{0}\equiv n_{c}(\dot\gamma=0)$. This figures presents our results as well as results of flow-induced nucleation in polymer melts taken from Refs.~\cite{Olmsted_PRL_103_2009,Graham_Olmsted_2010}. Remarkably, all the data collapse into a master-curve reproducible by
\begin{equation}
\chi(n_{c})=\frac{n_{c}-n_{0}}{n_{c}+n_{0}}.\label{eq_chi_nc_fit}
\end{equation}
The Eq.~(\ref{eq_chi_nc_fit}) follows directly from Eqs.~(\ref{eq_nc3}) and (\ref{eq_chi_tau}).

As seen from Fig.~\ref{fig_10}(b), values of $n_c/n_0$ and $\chi$ accessible to the considered Dz-system correspond to a narrower range in comparison with values for the polymer system. The Dzugutov interaction potential is not capable of forming elongated structures with large values of deformation parameter. Therefore, it is expected that the results for the Dzugutov pseudo-metallic system are located in the lower left part of the figure, whereas the data for the soft systems (say, polymers) have to be located at higher values of parameters. Results presented in Fig.~\ref{fig_10}(b) indicate on a possible universality in correlation between the characteristics of size and shape critically-sized nucleus, that does not depend on specific types of the considered systems. Therefore, additional studies of this point would be useful.

\section{Conclusion}

The main results of this study are the following:

(i) The shear-induced structural ordering of the single-component glassy system is initiated through crystal nucleation mechanism.

(ii) The critical size $n_{c}$ and the nucleation time $\tau_{c}$ were evaluated at different fixed shear rates and temperatures. It is shown that both the size and the shape of the critically-sized nuclei depend on the shear rate. The asphericity of the nuclei increases with the increasing shear rate $\dot\gamma$ (or with the increasing critical deformation $\gamma_{c}\equiv\dot\gamma\tau_{c}$); the larger asphericity of the nucleus corresponds to the larger critical size. It is verified by the increasing asphericity parameter $S_{0}$ and by the changes of the contours of $g_{c}(x,y)$ and $g_{c}(x,z)$ calculated for the particles of the critically-sized nuclei.

(iii) The shapes of the critically-sized nuclei change with increasing shear rate from spherical to ellipsoidal. The asphericity parameter and the critical size increase according to the power-law $\propto(\dot\gamma\tau_{c})^{1/3}$.

(iv) It is shown that the ellipsoidal nuclei are oriented within the shear-gradient plane at moderate and high shear rates. The tilt angle of the nuclei increases with the increasing shear rate according to power-law dependence.

(v) It is found that the scaled $n_{c}$-dependencies of the nuclei deformation parameter $\chi$ are collapsed into unified master-curve.

\section*{Acknowledgments}
The work is supported in part by the grant MD-5792.2016.2 (program of support for young scientists in RF). Authors are grateful to the Ministry of Education and Science of the Russian Federation for supporting the research in the framework of the state assignment (No 3.2166.2017/4.6). The molecular dynamic simulations were performed by using the computational clusters of Kazan Federal University and of Joint Supercomputer Center of RAS.

\bibliographystyle{unsrt}

\end{document}